\documentclass[onefignum,onetabnum,onealgnum]{siamonline190516}

\usepackage[dvipsnames]{xcolor}

\usepackage{comment}
\usepackage{lipsum}
\usepackage[margin=1in, footskip=36pt]{geometry}

\usepackage{endnotes}

\usepackage{titlesec}
\usepackage{titletoc}

\titleclass{\task}{straight}[\section]
\newcounter{task}
\renewcommand{\thetask}{\arabic{task}}
\titleformat{\task}[hang]
    {\normalfont\LARGE\bfseries}{Task \thetask:}{1em}{}

\titleformat*{\task}{\color{header1}\bfseries}

\titlecontents{task}
              [3.8em] 
              {}
              {\contentslabel{2.3em}}
              {\hspace*{-2.3em}}
              {\titlerule*[1pc]{.}\contentspage}

\titlespacing*{\section}{0ex}{1ex}{1ex}
\titlespacing*{\subsection}{0ex}{1ex}{1ex}
\titlespacing*{\paragraph}{0ex}{1ex}{1ex}
\titlespacing*{\subparagraph}{0pt}{1ex}{1ex}
\titlespacing*{\task}{0em}{1ex}{1ex}

\usepackage{amsfonts,amsmath,amssymb}
\usepackage{bbm}

\usepackage{multicol}
\usepackage{enumitem}
\setlist[enumerate]{wide, labelindent=1cm,  noitemsep}
\setlist[itemize]{noitemsep}
\setlist[description]{noitemsep}

\usepackage{hhline}

\usepackage{todonotes}
\RequirePackage{ulem}

\usepackage{booktabs}
\usepackage{titlesec}
\usepackage{fancyhdr}
\usepackage{fullpage}

\RequirePackage{titlesec}
\RequirePackage[font={small},{singlelinecheck=false}]{caption}
\usepackage[T1]{fontenc}
\usepackage[scaled]{helvet}
\usepackage[scaled=1.10, med]{zlmtt}

\usepackage{algorithm}
\usepackage{algpseudocode}

\providecommand{\mb}[1]{\boldsymbol{#1}}

\usepackage[sort&compress,comma,square,numbers]{natbib}

\usepackage{color}
\usepackage{pifont}
\definecolor{aogr}{rgb}{0.0, 0.5, 0.0}

\title{
Valid Two-Sample Graph Testing via Optimal Transport Procrustes and Multiscale Graph Correlation with Applications in Connectomics
}

\author{
Jaewon Chung$^{a,\dagger,}${\thanks{Corresponding author: \email{j1c@jhu.edu};
$^\dagger$These authors contributed equally\newline
$^a$Johns Hopkins University, Baltimore, MD 21218, United States\newline
$^b$Texas A\&M, College Station, TX 77843, United States\newline
$^c$North Carolina State University, Raleigh, NC 27695, United States}
},
Bijan~Varjavand$^{a,\dagger}$,
Jes\'us Arroyo$^{b}$,
Anton Alyakin$^a$,
Joshua Agterberg$^a$,
Minh Tang$^c$,
Carey E.~Priebe$^a$,
Joshua T.~Vogelstein$^{a}$
}

\begin{document}

\maketitle

\begin{abstract}
Testing whether two graphs come from the same distribution is of interest in many real world scenarios, including brain network analysis.
Under the random dot product graph model, the nonparametric hypothesis testing framework consists of embedding the graphs using the adjacency spectral embedding (ASE), followed by  aligning the embeddings using the median flip heuristic, and finally applying the nonparametric maximum mean discrepancy (MMD) test to obtain a p-value.
Using synthetic data generated from \textit{Drosophila} brain networks, we show that the median flip heuristic results in an invalid test, and demonstrate that optimal transport Procrustes (OTP) for alignment resolves the invalidity.
We further demonstrate that substituting the MMD test with multiscale graph correlation (MGC) test leads to a more powerful test both in synthetic and in simulated data.
Lastly, we apply this powerful test to the right and left hemispheres of the larval \textit{Drosophila} mushroom body brain networks, and conclude that there is not sufficient evidence to reject the null hypothesis that the two hemispheres are equally distributed.
\end{abstract}
\noindent%

\section{Introduction}\label{sec:introduction}
A network, or graph, is a data type which naturally encodes information about relationships between variables.
Statistical network analysis is becoming an increasingly important area \cite{goldenberg2010survey}, as it has applications in fields such as brain \cite{bullmore2009complex} and social sciences \cite{wasserman1994social}. Often, we encounter more than one graph observation, and it is  scientifically interesting to determine whether the two graphs come from the same distribution: the \textit{two-sample graph hypothesis testing} problem.
When the two samples are real-valued scalars, procedures such as the $t$-test and Wilcoxon rank-sum test are available, but these methods do not generalize to more complex data types such as networks.

Recently, methods have been proposed for determining whether two graphs are statistically equivalent under different settings \cite{semipar, omni, tang2014nonparametric, graph-comparison-same-1, graph-comparison-same-3, graph-comparison-same-4, graph-comparison-other-1, graph-comparison-other-2, graph-comparison-other-5, graph-comparison-other-8, graph-comparison-other-9}.
Most of these methodologies are aimed for pairs of graphs with a known correspondence between their vertices, and thus the problem consist in finding significant differences on the corresponding edges or vertices.
Here, we focus on the more general setting where this vertex correspondence is unknown or might not exist, for example when the graphs do not have the same number of vertices.
In this setting, we are interested in identifying significant differences on some underlying structure of the vertices.
In particular, \citet{tang2014nonparametric} proposed a nonparametric approach that operates on pairs of graphs in which there is no known correspondence between their corresponding vertex sets using the distance between probability distributions on a vertex latent space. This method uses the maximum mean discrepancy (MMD) test to detect differences between  distributions of the estimated latent positions of the vertices.
In \citet{agterberg2020nonparametric}, the authors extend this procedure to resolve the inherent non-identifiabilities in the vertex latent space via an optimal transport solution that yields consistent nonparametric hypothesis test.

MMD has been shown to be equivalent to the Energy distance two-sample test, the Hilbert-Schmidt independence
criterion (HSic), and the distance correlation test for independence (DCorr) \cite{exact-equivalence-1, exact-equivalence-2}.
HSic and DCorr, which are independence tests, are used as two-sample tests via first performing a \textit{k-sample transform}, which consists of concatenating the two samples, defining an auxiliary label vector, and subsequently testing for independence of the samples and label vector \cite{exact-equivalence-2}.
Multiscale graph correlation (MGC) is a recently proposed measure of dependence that
has shown an improved empirical power in many settings by intelligently selecting the appropriate
scale of the data \cite{mgc-0, mgc-1, mgc-2}. In this paper, we propose a new methodology for testing equivalence of distributions between networks using MGC as the test statistic. We demonstrate empirically in multiple simulation and synthetic data settings that
MGC outperforms other methods, and demonstrate its use on the problem of comparing the connectivity of the brain hemispheres of the \textit{Drosophila melanogaster}.

\section{Background}\label{sec:background}
\subsection{Graphs and Embeddings}
A graph $G = (\mathbb{V}, \mathbb{E})$ with $n$ vertices is composed of a vertex set $\mathbb{V} = \{v_1, \dots, v_n\}$ and an edge set $\mathbb{E} \subseteq \mathbb{V} \times \mathbb{V}$, where the edge $(v_i,v_j)\in\mathbb{E}$ connects vertices $i$ and $j$. Graphs can be represented by an adjacency matrix $A\in\{0,1\}^{n\times n}$, with rows and columns corresponding to vertices and matrix entries corresponding to edge values, so $A_{ij}=1$ whenever $(v_i,v_j)\in\mathbb{E}$.

The \textit{random dot product graph} (RDPG) model \cite{athreya2018rdpg} treats the entries of an adjacency matrix $A$ as independent Bernoulli random variables, where the probability of an edge is given by the dot product of pairs of latent positions $x_1,\ldots, x_n\in\mathbb{R}^d$ for each vertex, so $\mathbb{P}(A_{ij}=1) = x_i^\top x_j$. These latent positions are independent random variables sampled according to some distribution $F$. Writing $X = \left[x_1 \cdots x_n\right]^\top$ as the matrix of latent positions, we denote $(X,A)\sim\text{RDPG}_n(F)$ as a RDPG with adjacency matrix $A$ and (usually unobserved) latent positions $X$ sampled from $F$. With this notation, we have that $\mathbb{E}(A|X) = XX^{\top}$.

The RDPG model provides a flexible framework for studying the statistical equivalence of a pair of graphs. Suppose that $(X,A)\sim\text{RDPG}_n(F_X)$ and $(Y,B)\sim\text{RDPG}_m(F_Y)$.
Then, the two graphs $A$ and $B$ are said to have the same distribution if there exists an orthogonal matrix $W\in\mathbb{R}^{d\times d}$, $W^\top W = I$ that makes $X$ and $W^\top Y$ have the same distribution.
The matrix $W$ accounts for the nonidentifiability inherent with inner products \cite{tang2014nonparametric}.
Formally, this hypotheses test can be stated as
  \begin{align*}
    &H_0 : F_X = F_Y \circ {W}
    && \text{for some orthogonal operator } {W},\\
       &H_A : F_X \neq F_Y \circ {W}
    && \text{for all orthogonal operators } {W}.
  \end{align*}
Here, $F_Y\circ W$ denotes the distribution of the random variable $W^\top Y$.
The graphs $A$ and $B$ do not need to have a correspondence between their vertices, or even
the same number of vertices, because we are comparing distributions of latent
positions instead of the latent positions themselves.
This work focuses on cases where the number of vertices are equal, or approximately equal, that is $n \approx m$.

Latent positions are typically unobserved in practice and can be estimated via the \textit{adjacency spectral embedding} (ASE) \cite{ase-consistency-1}.
Suppose that $A= USV^\top + U_\perp S_\perp V^\top_\perp$ is the singular value decomposition of $A$, where $U,V\in\mathbb{R}^{n\times d}$, $U_\perp,V_\perp\in\mathbb{R}^{n\times(n-d)}$ are jointly orthogonal matrices
corresponding to the singular vectors of $A$, and $S\in\mathbb{R}^{d\times d}$,
$S_\perp\in\mathbb{R}^{(n-d)\times (n-d)}$ are diagonal matrices such that $S$
contains the $d$ largest singular values of $A$.
Then, the ASE of $A$ is defined as $\hat{X}=US^{1/2}$. This simple and computationally efficient approach results in consistent estimates $\hat{X}$ and $\hat{Y}$ of the true latent positions $X$ and $Y$ \cite{ase-consistency-1,ase-consistency-2, ase-consistency-3}.
The ASE depends on a parameter $d$ that corresponds to the rank of the expected adjacency matrix conditional on the latent positions; in practice, we estimate this dimension, $\hat d$ via the scree plot of the eigenvalues of the adjacency matrix which can be done automatically via a likelihood profile approach \cite{zhu2006automatic}.

If the two networks have large difference in the number of vertices, the subsequent testing procedure might be invalid due to the finite-sample variances of the estimated latent positions, $\hat{X}$ and $\hat Y$, that depend on the number of vertices \cite{Athreya2015}. Thus, the distributions of the estimated latent positions may not be the same even if the true distributions are the same. The difference in variances can be corrected by adding scaled Gaussian noise to the estimated latent positions of larger network, which increases the variance of the larger network to be approximately the same as that of the smaller network \cite{correcting-nonpar}. This correction results in a valid test for the equivalence of the distributions of latent positions.

\subsection{Orthogonal Nonidentifiabilities}
There are two sources of an orthogonal nonidentifiability associated with using the
ASE in RDPG \cite{on-two-sources}. The first is associated with the RDPG model itself, and can take a form of any orthogonal transformation, since for any
orthogonal matrix $W$, it holds that $(XW)(XW)^\top = X W W^\top X^\top = X X^\top$. When
using ASE, this orthogonal matrix converges to a population value at the rate
$O_p \left(n^{-1/2}\right)$, and, thus, rarely has any inferential consequences \cite{tang2014nonparametric, on-two-sources}.

The second source, called subspace nonidentifiability, is associated with taking a singular value decomposition as a part of the ASE.
Consider the singular value decomposition of the matrix $XX^\top$.
Since it is positive semidefinite, its eigendecomposition and singular value decomposition coincide.
If $XX^\top$ has no repeated singular (eigen) values, then each singular (eigen) vector is determined only up to a sign ambiguity.
However, if $XX^\top$ has repeated eigenvalues, then the singular (eigen) vectors corresponding to each repeated singular value are only unique up to an orthogonal transformation in the dimension of the corresponding subspace.
Since one only observes two different adjacency matrices, then the leading singular vectors may not be aligned \textit{a priori}.
If one assumes that $XX^\top$ has distinct eigenvalues, this sign ambiguity is often adjusted for by flipping the signs of each dimensions such that the medians have the same sign; that is for each embedding dimension, $\text{sign}(\text{median}(X_i)) = \text{sign}(\text{median}(Y_i))$ for all $i\in d$ \cite{tang2014nonparametric, correcting-nonpar}.
This heuristic can perform poorly when the medians of the samples are too close to zero.
In addition, this approach falls short in the case of repeated eigenvalues of the matrix $XX^{\top}$ since the subspace associated with such values are only unique up to a more general rotation, and hence the leading singular vectors of each adjacency matrix may not be close, even if $n$ is large, since the leading singular values of the adjacency matrices will be perturbed versions of the singular values of $XX^\top$. For more details and discussion on this form of nonidentifiability, see \citet{on-two-sources}.

If the two graphs had paired vertices, then the orthogonal misalignment between the two samples could be resolved by using a solution to
a well-known orthogonal Procrustes problem \cite{schonemann1966generalized}.
In unpaired graphs, there is no \textit{a priori} assignment between vertices, so we can employ an Optimal Transport Procrustes (OTP) algorithm \cite{agterberg2020nonparametric}, which simultaneously solves the alignment and the assignment problems. Formally, this algorithm minimizes the objective function
\begin{align}
    \min_{W,  \Pi} \langle \Pi, C_W \rangle
    \label{eq:2.1}
\end{align}
where the entries of the cost matrix $C_W$ are given by $\left(C_W\right)_{ij} =  ||\hat X_i - W\hat Y_j||^2$,  $\hat X \in \mathbb{R}^{n\times d}$ and $\hat Y \in \mathbb{R}^{m\times d}$ are estimated latent positions,  $\Pi = \frac{1}{nm} \mb{1} \mb{1}^\top$ is an assignment matrix that satisfies $\Pi \mb{1} = \frac{1}{n} \mb{1}$ and $\Pi^{\top} \mb{1} = \frac{1}{m} \mb{1}$, and $W$ is constrained to be orthogonal. Given an initial guess $W_0$, the algorithm iteratively updates $\Pi_{i+1} | W_{i}$
by via an optimal transport algorithm \cite{alvarez2019towards} and $W_{i+1} | \Pi_{i}$ using the solution to the Procrustes problem. In graph setting, the algorithm is initialized with all possible orthogonal diagonal matrices, i.e. the
$2^d$ different diagonal matrices with $\pm{1}$ on the diagonal \cite{agterberg2020nonparametric}. In \citet{agterberg2020nonparametric}, the authors show that the orthogonal matrix that globally minimizes the objective function \eqref{eq:2.1} yields a consistent estimate of the orthogonal matrix approximately aligning the empirical distributions of the two ASEs, including when the corresponding graphs have (asymptotically) repeated singular (eigen) values.

\subsection{Distance Correlation}
Due to the equivalence between two-sample and independence testing \cite{exact-equivalence-2}, distance correlations (DCorr) can be used to test the equality of the distributions $F_X$ and $F_Y$. Define $Z=(X^\top, Y^\top)^\top\in\mathbb{R}^{N\times d}$ and $E=(0_n, 1_m)^\top\in\mathbb{R}^{N}$, where $N=n+m$. DCorr tests the independence of $Z$ and $E$ using some distance functions $\delta_Z:\mathbb{R}^{d}\times\mathbb{R}^{d}\rightarrow\mathbb{R}$ and $\delta_E:\mathbb{R}\times\mathbb{R}\rightarrow\mathbb{R}$. Here, $\delta_Z$ and $\delta_E$ denote the Euclidean norm in $\mathbb{R}^{d}$ and $\mathbb{R}$, respectively.

First, DCorr computes distance matrices $D^Z, D^E$ such that $D^Z_{i,j} = \delta_Z(Z_i, Z_j)$ and $D^E_{i,j} = \delta_E(E_i, E_j)$. The distance matrices are then doubly centered to $D^{Z'}, D^{E'}$, where $D^{Z'}_{i,j} = D^Z_{i,j} - \overline{D^Z}_{\cdot,j} - \overline{D^Z}_{i,\cdot} + \overline{D^Z}_{\cdot,\cdot}$, and similarly for $D^{E'}$. Here the column means, row means, and the grand mean are $\overline{D^Z}_{\cdot, j} = \frac{1}{N}\sum_{i=1}^N D^Z_{i,j}$, $\overline{D^Z}_{i, \cdot} = \frac{1}{N}\sum_{j=1}^N D^Z_{ij}$, and $\overline{D^Z}_{\cdot, \cdot} = \frac{1}{N^2}\sum_{i=1}^N\sum_{j=1}^N D^Z_{ij}$, respectively. The sample DCorr test statistic \cite{szekely2014dcorr} is defined as
\begin{align*}
    \operatorname{DCorr}(Z,E) = \frac{1}{N(N-3)\sigma_{D^{Z'}}\sigma_{D^{E'}}}\sum_{i,j} D^{Z'}_{i,j}D^{E'}_{i,j},
\end{align*}
where $\sigma_{D^{Z'}}$ and $\sigma_{D^{E'}}$ are the standard deviation of values in $D^{Z'}$ and $D^{E'}$, respectively. A null distribution of this test statistic
can be generated by permuting the indices of $E$ or approximated using an asymptotic result \cite{shen2019chi}.

The centered versions $D^{Z'}, D^{E'}$ have the property that all rows and columns sum to zero, but the test statistic is biased for finite samples. An unbiased version of DCorr modifies the centering method of the pairwise distance matrices \cite{szekely2014dcorr}. This method, called U-centering, generates a matrix $D^{Z''}$ that has the additional property that $E[D^{Z''}_{ij}] = 0, i, j = 1, \dots, N$. $D^{Z''}$ uses a slightly different form for the row means, column means, and grand mean, given by $\widetilde{D^{Z}}_{\cdot, j} = \frac{1}{N-2}\sum_{i=1}^N D^{Z}_{ij}$ (similarly for row means) and $\widetilde{D^{Z}}_{\cdot, \cdot} = \frac{1}{(N-2)(N-1)}\sum_{i,j} D^{Z}_{ij}$. The test statistic is defined analogously modulo these new definitions.

\subsection{Multiscale Graph Correlation}
An alternative method for approaching the hypothesis testing problem is MGC \cite{mgc-0, mgc-1, mgc-2}, which uses the distance-based methods in DCorr, but also considers the local scale of the data.
MGC is based on unbiased DCorr, resulting in it also being unbiased. MGC consists of the following steps.

\begin{enumerate}
	\item Compute the centered distance matrices $D^{Z''}$ and $D^{E''}$.
	\item For each $k,l$ in $1,\ldots, N$, compute the $k$ and $l$ nearest neighbors of each row of $D^{Z''}$ and $D^{E''}$, respectively, and denote the induced nearest neighbor graphs by $G^k\in\{0,1\}^{N\times N}$ and $H^l\in\{0,1\}^{N\times N}$.
	\item Estimate the local normalized correlations $c^{k,l}$ such that
	\begin{equation*}
	c^{kl} = \frac{\sum_{i,j}D^{Z''}_{ij}D^{E''}_{ij}G^k_{ij}H^{l}_{ij}}{\left(\sum_{i,j}(D^{Z''}_{ij})^2G^k_{ij}\right) \left(\sum_{ij}(D^{E''}_{ij})^2H^{l}_{ij}\right)}.
	\end{equation*}
	\item Using a smoothing parameter $\tau$, estimate the smoothed maximum of $c^{kl}$ over all possible values of $k$ and $l$, defined as
	\begin{align*}
	R &= \text{LCC}\{(k,l):c^{kl}>\max(\tau,c^{NN})\},\\
    c^\ast &= \max_{(k,l)\in R}c^{kl},
	\end{align*}
	where LCC denotes the largest connected component of the graph defined by a set of edges.
\end{enumerate}

Similar to DCorr, the null distribution of the test statistic, $c^\ast$, is generated by permuting the indices of $E$.
When the relationship is nonlinear or non-monotonic, MGC tends to choose scales smaller than $n$, detecting relationships more often than DCorr, and thus, it can be considered a direct generalization to the above methods, with often large finite sample power gains, with a minor running time cost \cite{mgc-0}.

\section{Simulations}\label{sec:simulations}

The performance of DCorr and MGC is analyzed by simulating graphs with different distributions. The graphs are simulated according to the RDPG model, using different distributions (described below) to generate their univariate latent positions $X=(x_1,\ldots, x_n)^\top$ and $Y=(y_1,\ldots, y_n)^\top$, and the latent positions are estimated via ASE with embedding dimension $\hat d= 1$.
The sign ambiguity is resolved via median flip since the graphs are embedded into one dimension. The estimates are then used to compute the corresponding test statistics, and calculate the $p$-value after estimating the null distribution via permutation test. The tests reject at a significance level $\alpha=0.05$, and the empirical power is based on 1000 Monte Carlo replicates. This process is repeated for increasing numbers of vertices in each graph. The graph generating mechanisms are described next.

\begin{figure}[hbt!]
    \centering
    \includegraphics[width=\linewidth]{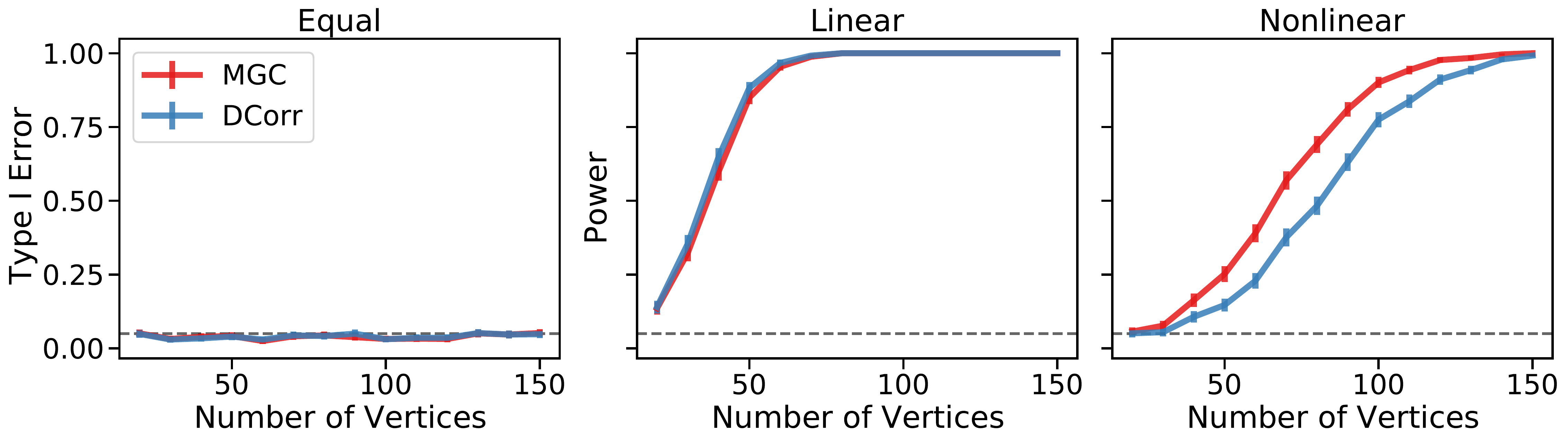}
    \caption{Type I error and power vs. number of vertices ($n$) for detecting differences in the distribution of latent positions using different methods (DCorr and MGC). Dashed lines represent $\alpha=0.05$, and error bars represent $95\%$ confidence interval.
    \textit{(Left)} When the distributions of the latent positions are identical, both methods are valid as power is near or at $\alpha=0.05$. \textit{(Middle)} When the distributions are linearly dependent, both methods are equally powerful and consistent. \textit{(Right)} When the distributions are nonlinearly dependent, both methods are consistent, but MGC is more powerful than DCorr.}
    \label{fig:simulations}
\end{figure}

\paragraph{Equal distribution of the latent positions}
First, we analyze the performance of the methods when the null hypothesis is true, so the latent positions $X$ and $Y$ have the same distribution: $x_i \overset{\text{iid}}{\sim} F_X$ and $y_i \overset{\text{iid}}{\sim} F_X$, $i=1,\ldots,n$,
with
$F_X = \operatorname{Unif}(.2,.7)$.
That is, these mutually independent sets of latent positions are uniformly distributed on the range $(.2, .7)$.
Figure \ref{fig:simulations} (left) shows the empirical size of the tests for different numbers of vertices. The empirical power does not exceed $\alpha$, showing that both DCorr and MGC correctly control the type I error.

\paragraph{Linear difference in the distributions}
For this setting, we generate $x_i \overset{\text{iid}}{\sim} F_X$ and $y_i \overset{\text{iid}}{\sim} F_X + 0.1$ (Figure \ref{fig:simulations}, middle).
As the number of vertices in each graph increases, the difference between the distributions becomes easier to detect, and both DCorr and MGC algorithms detect the differences, resulting in power increasing to unity at equal rate for both methods.

\paragraph{Nonlinear difference in the distributions}
Finally, set $x_i \overset{\text{iid}}{\sim} F_X$ and $y_i \overset{\text{iid}}{\sim} 0.5\operatorname{Beta}(.2, .2)+0.2$. Figure \ref{fig:simulations} (right) shows  the power of both DCorr and MGC goes to 1, but MGC dominates DCorr at  all sample sizes sample sizes until both tests achieve power of one.

\section{Synthetic and Real Data Applications}

\paragraph{\textit{Drosophila melanogaster} brain}
The connectomes, or brain graphs, of the \textit{Drosophila melanogaster} mushroom body was obtained in \citet{eichler2017complete}. The left brain graph ($A^L$) has $n=163$ neurons represented with the vertices, and the right brain graph ($A^R$) has $m=158$ neurons. Both graphs are binary with $A_{i j} \in \{0, 1\}$, symmetric with $A= A^\top$, and hollow with $\text{diag}(A) = \vec 0$. The edges of the data represent whether a synaptic connection exist between a pair of neurons or not.

\paragraph{Synthetic Data Analysis}
The performances of the two SVD alignment methods (median flipping and Optimal Transport Procrustes) and two hypothesis tests (DCorr and MGC) are analyzed by simulating synthetic data from the \textit{Drosophila} connectomes. Given an estimated latent position matrix $\hat X\in \mathbb{R}^{n\times d}$ obtained from one of the hemispheres (either $A^L$ or $A^R$), a pair of new latent position matrices are sampled, either with or without perturbation, to analyze the power and validity of various methods as a function of the number of vertices and the effect size. The latent position matrices, $\hat X$, are multivariate with $d=3$ for both hemispheres, which is given by the likelihood profile of eigenvalues \cite{zhu2006automatic}.

The first step in data generation is to sample with replacement a set of $m$ vertices from the original graph. Let $\mathcal{P}\subseteq [n]$ be the set of indices for the randomly selected vertices, where $|\mathcal{P}|=m$. For each $i\in\mathcal{P}$, a pair of $d$-dimensional random latent positions $Y_i$ and $Z_i$ are independently generated
via
\[Y_i \sim \mathcal{MVN}(\hat X_i, \hat\Sigma(\hat X_i)), \quad\quad \quad Z_i \sim  \mathcal{MVN}(\hat X_i + \epsilon_i, \hat\Sigma(\hat X_i)),\]
where $\mathcal{MVN}(x, \Sigma)$  is a $d$-variate normal distribution with mean $x$ and covariance $\Sigma$, and $\hat\Sigma(\hat X_i)$ is the covariance matrix of the difference between $\hat{X}_i$ and  $X_i$ given by the central limit theorem in \citet{Athreya2015}, which is introduced to account for the fact that the true latent position $X_i$ is not observed. A random subset $\mathcal{O} \subseteq \mathcal{P}$ of the vertices are perturbed by adding $\epsilon_i$,  defined as
\begin{align*}
\epsilon_j &\sim
\begin{cases}
     \text{Unif}(\mathcal{S}^d_r), & \text{if}~j \in \mathcal{O}\\
    \vec 0, & \text{otherwise},\\
\end{cases}
\end{align*}
where $\text{Unif}(\mathcal{S}^d_r)$ is the uniform distribution on the surface of a $d$ dimensional sphere with radius $r$. Thus, $r$ represents the effect size, and $\rho = |\mathcal{O}| / m$ is the proportion of changed vertices. Given the new latent position matrices $Y$ and $Z$, a pair of undirected graphs $A$ and $B$ are generated by sampling the edges independently as $A_{ij}\sim\text{Bernoulli}(\min\{1, \max\{0, Y_i^\top Y_j\}\})$ and $B_{ij}\sim\text{Bernoulli}(\min\{1, \max\{0,Z_i^\top Z_j\}\})$, for $i>j$. $A$ and $B$ are symmetrized by setting $A_{ij} = A_{ji}$ and $B_{ij} = B_{ji}$.

\begin{figure}[b!]
    \centering
    \includegraphics[width=1\linewidth]{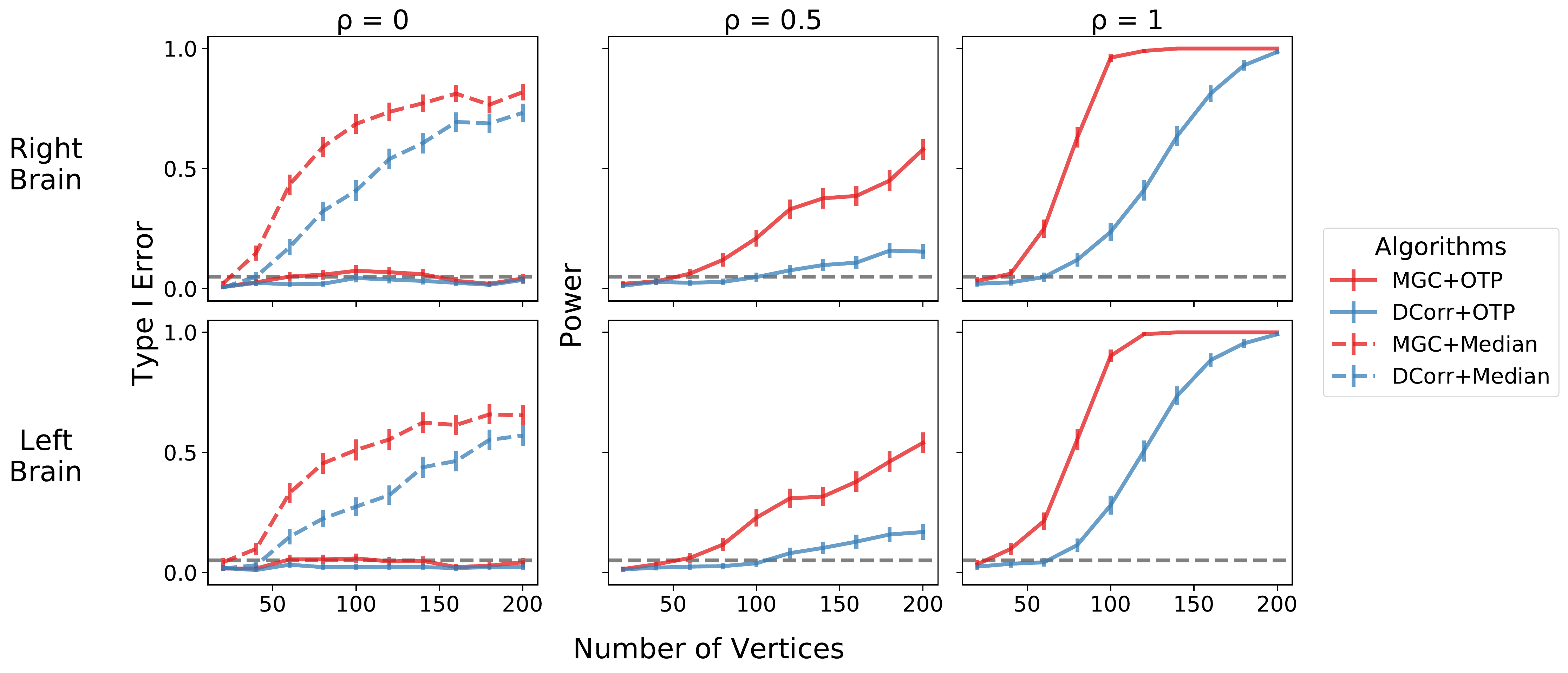}
    \caption{
    Type I error and power vs. number of vertices ($n$) using synthetic data generated from \textit{Drosophila} connectomes. The effect size, or the magnitude of the change in latent positions, is fixed at $r = 1$. The gray dashed line is $\alpha=0.05$.
    Each row represents right or left hemisphere of the brain, and each column represents different proportion of vertices changed. When $\rho = 0$, the latent position distributions $F_V$ and $F_W$ are the same. Latent position alignment via median flip is invalid as shown in first column ($\rho = 0$) as type I error is above $\alpha$ and increasing with the size of the graphs; OTP resolves the invalidity of median flip.
    Testing via MGC is more powerful than testing via DCorr as shown in columns where $\rho = 0.5$ and $\rho = 1$.
    Curves for median flip testing are omitted in the middle and right panels since using median flip results in an invalid test.
    }
    \label{fig:synthetic_power}
\end{figure}

Once the new adjacency matrices $A$ and $B$ are sampled, latent positions are estimated with embedding dimension $\hat{d} = 3$, and then we test the hypothesis that the distributions used to generate their latent positions are the same. 
The hypotheses are rejected at a significance level $\alpha=0.05$, and the empirical power based on 500 Monte Carlo replicates is reported. This process is repeated for increasing number of vertices with $m\in[20, 200]$, and proportion of changed vertices with $\rho \in \{0, 0.5, 1\}$, while keeping the effect size constant with $r = 1$ for both left and right \textit{Drosophila} mushroom body. The effect size is fixed because there is not enough influence on the power at smaller values of $r$.

Figure \ref{fig:synthetic_power} shows that aligning the latent positions via the median flip yields invalid results for both DCorr and MGC.
Specifically, when $\rho=0$, both $Y$ and $Z$ are sampled from the same distribution, but both DCorr and MGC yield type I errors  greater than $0.05$ when testing via median flip.
The OTP alignment resolves the invalidity of the median flip as type I error is near $\alpha = 0.05$ at all sample sizes.
When $\rho>0$, the empirical power for both DCorr and MGC using OTP increases as the the sample size and proportion of vertices changed increases, showing that both methods are able to identify a significant difference between the distributions.
Lastly, the MGC test statistic is more powerful than the DCorr test statistic in all scenarios where $\rho >0$.

\paragraph{Left vs Right Larval \textit{Drosophila} Mushroom Body}

The left and right \textit{Drosophila} mushroom body are similar, but not identical. To compare  these hemispheres, we test the difference of the distributions of the graph of the left mushroom body ($L$) and the graph of the right mushroom body ($R$).
While the left and right mushroom body connectomes have different number of vertices, we did not correct the embeddings since the difference is small.
The ASE of left and right hemispheres are denoted by $\hat{X}_L$ and $\hat{X}_R$ with assumed distributions $F_L$ and $F_R$, respectively.
We test the hypothesis $H_0: F_L = F_R$ vs $H_1: F_L \neq F_R$ with various embedding dimensions, $\hat d \in \{1, 2, 3, 4, 5\}$. The $p$-values are shown in Table \ref{tab:pvals}.

\begin{figure}[t!]
    \centering
    \includegraphics[width=\linewidth]{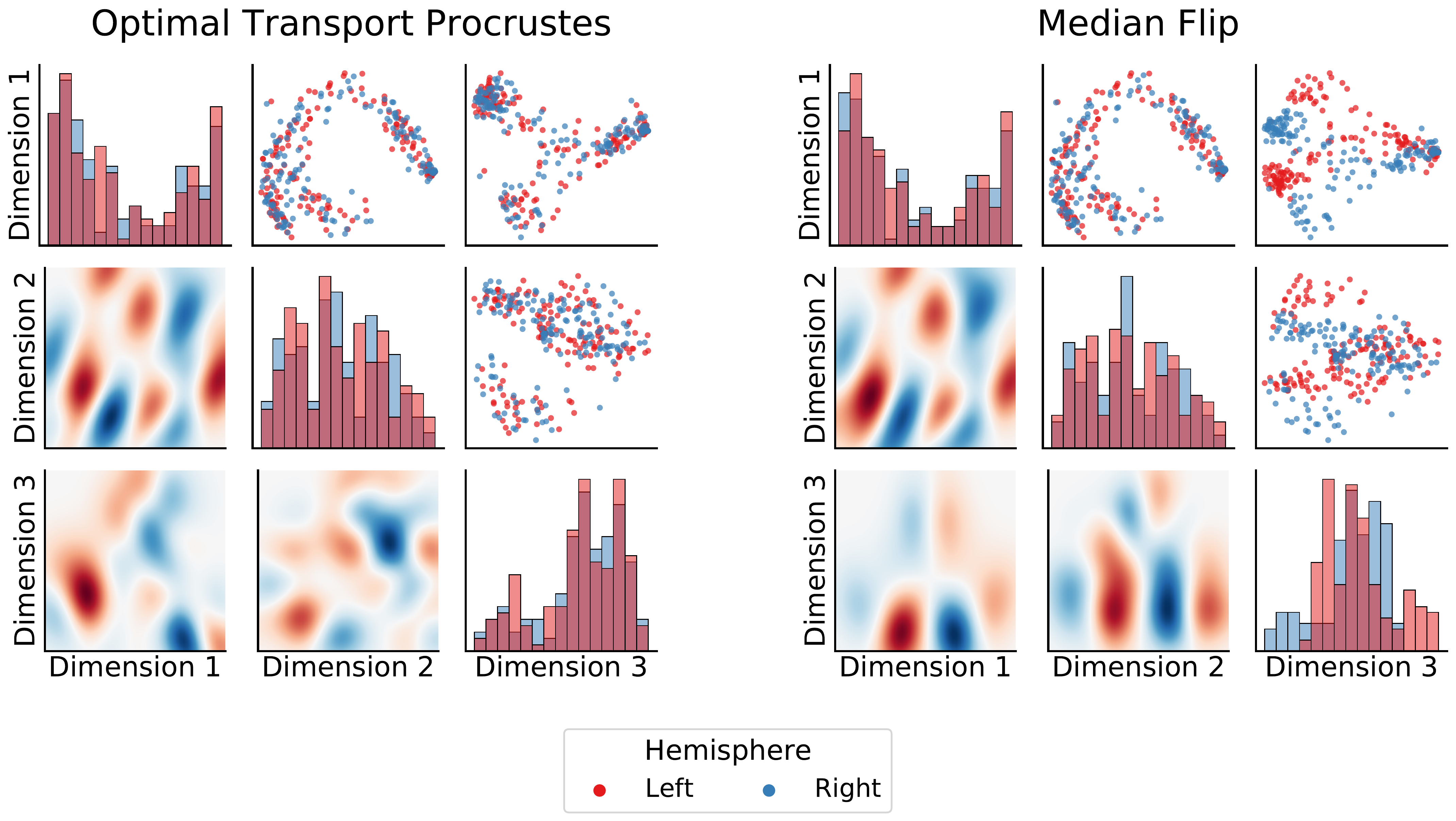}
    \caption{Embeddings for the \textit{Drosophila} connectomes after alignment with Optimal Transport Procrustes (left) and median flip (right) with embedding dimension $\hat d =3$. The diagonals are histograms of each dimension, upper off-diagonals are pairwise scatter plots, and lower off-diagonals are the difference of the kernel density estimates of the two embeddings.
    \textit{(Left)} OTP properly aligns all three dimensions.
    \textit{(Right)} Median flips results in misalignment in dimension 3 of the left hemisphere embedding.
    }
    \label{fig:drosophila}
\end{figure}

At lower embedding dimensions ($\hat d \in \{1, 2\}$), testing via median flip and OPT for MGC and DCorr do not reject the null.
However, at higher embedding dimensions ($\hat d \in \{3, 4, 5\}$), testing via median flip rejects the null, but testing via OPT does not for both MGC and DCorr.
Figure \ref{fig:drosophila} shows that the estimated latent positions of both hemisphere are similar, suggesting there is no difference in the distributions, but median flip misaligns the third dimension.
The misalignment causes MGC and DCorr to reject the null.
Thus, the left and right \textit{Drosophila} mushroom body are not significantly different.

\begin{table}
\centering
\begin{tabular}{llllll}
\toprule
\textbf{Algorithm} &       \multicolumn{5}{c}{\textbf{$p$-value}}      \\
\cmidrule(lr){2-6}
  &      $\hat d = 1$ &    $\hat d = 2$ &  $\hat d = 3$ & $\hat d = 4$ & $\hat d = 5$ \\
\midrule
MGC+OPT      &  \textcolor{ForestGreen}{0.986} &  \textcolor{ForestGreen}{1} &  \textcolor{ForestGreen}{1} &  \textcolor{ForestGreen}{0.999} &  \textcolor{ForestGreen}{0.952} \\
DCorr+OPT    &  \textcolor{ForestGreen}{0.985} &  \textcolor{ForestGreen}{1} &  \textcolor{ForestGreen}{0.997} &  \textcolor{ForestGreen}{0.998} &  \textcolor{ForestGreen}{0.951} \\
MGC+Median   &  \textcolor{ForestGreen}{0.993} &  \textcolor{ForestGreen}{1} &  \textcolor{Red}{0.001} &  \textcolor{Red}{0.001} &  \textcolor{Red}{0.001} \\
DCorr+Median &  \textcolor{ForestGreen}{0.986} &  \textcolor{ForestGreen}{1} &  \textcolor{Red}{0.005} &  \textcolor{Red}{0.007} &  \textcolor{Red}{0.038} \\
\bottomrule
\end{tabular}
\caption{P-values from testing for differences in distribution of the left and right \textit{Drosophila} mushroom body, specifically $H_0: F_L = F_R$ vs $H_1: F_L \neq F_R$ at various values of embedding dimension, $\hat d$. With $\alpha=0.05$, testing via optimal transport Procrustes and median flip do not reject the null hypothesis when $\hat d \in \{1, 2\}$. However, when $\hat d \geq 3$, testing via optimal transport Procrustes does not reject the null hypothesis, but testing via median flip does reject the null hypothesis when it should not. The green $p$-values correspond to successful alignment of latent positions, while red $p$-values correspond to misalignment of latent positions.}
\label{tab:pvals}
\end{table}

\section{Discussion}

The results presented herein demonstrate the improvement of the nonparametric two-sample graph testing presented in \citet{tang2014nonparametric} by aligning estimated latent positions via Optimal Transport Procrustes (OTP) and testing via multiscale graph correlation. For applications in connectomics, we show that the orthogonal nonidentifiabilities that arise from adjacency spectral embedding can significantly impacts the results. Specifically, the synthetic data experiments show that median flip can invalidate DCorr and MGC, but OTP resolves the inadequacy of median flip. Both simulated and synthetic experiments show that MGC is more powerful than DCorr when testing via OTP.  Thus, testing via OTP and MGC is not only more powerful, but also more trustworthy. While this work focuses on cases where the two graphs have the same number of vertices, the testing procedure can be extended to cases where the two networks have different number of vertices by correcting the variance of the estimated latent positions \cite{correcting-nonpar}.

\section{Code}
All code and data used in the analysis are available at \href{https://github.com/neurodata/improving-latent-distribution-test}{https://github.com/neurodata/improving-latent-distribution-test}.
The analysis were performed using the \texttt{graspologic} (\href{https://github.com/microsoft/graspologic}{https://github.com/microsoft/graspologic} \cite{chung2019graspy} and \texttt{hyppo}  \cite{panda2020hyppo} Python packages.
The methods described herein are implemented in \texttt{graspologic}.

\section*{Acknowledgment}
The authors would like to thank the NeuroData group for great advice and an equally great working environment. This material is based on research sponsored by the Defense Advanced Research Projects Agency (DARPA) Lifelong Learning Machines program [grant number FA8650-18-2-7834], the National Science Foundation award [grant number DMS-1921310], the Air Force Research Laboratory and DARPA [grant number FA8750-20-2-1001], and funding from Microsoft Research. The U.S. Government is authorized to reproduce and distribute reprints for Governmental purposes notwithstanding any copyright notation thereon. The views and conclusions contained herein are those of the authors and should not be interpreted as necessarily representing the official policies or endorsements, either expressed or implied, of the Air Force Research Laboratory and DARPA or the U.S. Government.

\clearpage
\bibliographystyle{unsrtnat}

\begin{thebibliography}{35}
\providecommand{\natexlab}[1]{#1}
\providecommand{\url}[1]{\texttt{#1}}
\expandafter\ifx\csname urlstyle\endcsname\relax
  \providecommand{\doi}[1]{doi: #1}\else
  \providecommand{\doi}{doi: \begingroup \urlstyle{rm}\Url}\fi

\bibitem[Goldenberg et~al.(2010)Goldenberg, Zheng, Fienberg, and
  Airoldi]{goldenberg2010survey}
Anna Goldenberg, Alice~X. Zheng, Stephen~E. Fienberg, and Edoardo~M. Airoldi.
\newblock A survey of statistical network models.
\newblock \emph{Found. Trends Mach. Learn.}, 2\penalty0 (2):\penalty0 129--233,
  February 2010.
\newblock ISSN 1935-8237.
\newblock \doi{10.1561/2200000005}.
\newblock URL \url{http://dx.doi.org/10.1561/2200000005}.

\bibitem[Bullmore and Sporns(2009)]{bullmore2009complex}
Ed~Bullmore and Olaf Sporns.
\newblock Complex brain networks: graph theoretical analysis of structural and
  functional systems.
\newblock \emph{Nature reviews neuroscience}, 10\penalty0 (3):\penalty0 186,
  2009.

\bibitem[Wasserman and Faust(1994)]{wasserman1994social}
Stanley Wasserman and Katherine Faust.
\newblock \emph{Social network analysis: Methods and applications}, volume~8.
\newblock Cambridge university press, 1994.

\bibitem[Tang et~al.(2017{\natexlab{a}})Tang, Athreya, Sussman, Lyzinski, Park,
  and Priebe]{semipar}
Minh Tang, Avanti Athreya, Daniel~L. Sussman, Vince Lyzinski, Youngser Park,
  and Carey~E. Priebe.
\newblock A semiparametric two-sample hypothesis testing problem for random
  graphs.
\newblock \emph{Journal of Computational and Graphical Statistics}, 26\penalty0
  (2):\penalty0 344--354, 2017{\natexlab{a}}.

\bibitem[{Levin} et~al.(2017){Levin}, {Athreya}, {Tang}, {Lyzinski}, and
  {Priebe}]{omni}
K.~{Levin}, A.~{Athreya}, M.~{Tang}, V.~{Lyzinski}, and C.~E. {Priebe}.
\newblock A central limit theorem for an omnibus embedding of multiple random
  dot product graphs.
\newblock In \emph{2017 IEEE International Conference on Data Mining Workshops
  (ICDMW)}, pages 964--967, 2017.
\newblock \doi{10.1109/ICDMW.2017.132}.

\bibitem[Tang et~al.(2017{\natexlab{b}})Tang, Athreya, Sussman, Lyzinski, and
  Priebe]{tang2014nonparametric}
Minh Tang, Avanti Athreya, Daniel~L. Sussman, Vince Lyzinski, and Carey~E.
  Priebe.
\newblock A nonparametric two-sample hypothesis testing problem for random
  graphs.
\newblock \emph{Bernoulli}, 23\penalty0 (3):\penalty0 1599--1630, 08
  2017{\natexlab{b}}.
\newblock \doi{10.3150/15-BEJ789}.
\newblock URL \url{https://doi.org/10.3150/15-BEJ789}.

\bibitem[Ghoshdastidar et~al.(2017)Ghoshdastidar, Gutzeit, Carpentier, and von
  Luxburg]{graph-comparison-same-1}
Debarghya Ghoshdastidar, Maurilio Gutzeit, Alexandra Carpentier, and Ulrike von
  Luxburg.
\newblock Two-sample tests for large random graphs using network statistics.
\newblock In Satyen Kale and Ohad Shamir, editors, \emph{Proceedings of the
  2017 Conference on Learning Theory}, volume~65 of \emph{Proceedings of
  Machine Learning Research}, pages 954--977, Amsterdam, Netherlands, 07--10
  Jul 2017. PMLR.
\newblock URL \url{http://proceedings.mlr.press/v65/ghoshdastidar17a.html}.

\bibitem[Levin and Levina(2019)]{graph-comparison-same-3}
Keith Levin and Elizaveta Levina.
\newblock Bootstrapping networks with latent space structure, 2019.
\newblock arXiv:1907.10821.

\bibitem[Arroyo et~al.(2019)Arroyo, Athreya, Cape, Chen, Priebe, and
  Vogelstein]{graph-comparison-same-4}
Jesús Arroyo, Avanti Athreya, Joshua Cape, Guodong Chen, Carey~E. Priebe, and
  Joshua~T. Vogelstein.
\newblock Inference for multiple heterogeneous networks with a common invariant
  subspace, 2019.

\bibitem[Rukhin and Priebe(2011)]{graph-comparison-other-1}
Andrey Rukhin and Carey~E. Priebe.
\newblock A comparative power analysis of the maximum degree and size
  invariants for random graph inference.
\newblock \emph{Journal of Statistical Planning and Inference}, 141\penalty0
  (2):\penalty0 1041 -- 1046, 2011.
\newblock ISSN 0378-3758.
\newblock \doi{https://doi.org/10.1016/j.jspi.2010.09.013}.
\newblock URL
  \url{http://www.sciencedirect.com/science/article/pii/S0378375810004246}.

\bibitem[Asta and Shalizi(2015)]{graph-comparison-other-2}
Dena~Marie Asta and Cosma~Rohilla Shalizi.
\newblock Geometric network comparisons.
\newblock In \emph{Proceedings of the Thirty-First Conference on Uncertainty in
  Artificial Intelligence}, UAI'15, pages 102--110, Arlington, Virginia, United
  States, 2015. AUAI Press.
\newblock ISBN 978-0-9966431-0-8.
\newblock URL \url{http://dl.acm.org/citation.cfm?id=3020847.3020859}.

\bibitem[Maugis et~al.(2020)Maugis, Olhede, Priebe, and
  Wolfe]{graph-comparison-other-5}
P.-A.~G. Maugis, S.~C. Olhede, C.~E. Priebe, and P.~J. Wolfe.
\newblock Testing for equivalence of network distribution using subgraph
  counts.
\newblock \emph{Journal of Computational and Graphical Statistics}, 0\penalty0
  (0):\penalty0 1--11, 2020.
\newblock \doi{10.1080/10618600.2020.1736085}.
\newblock URL \url{https://doi.org/10.1080/10618600.2020.1736085}.

\bibitem[Ghoshdastidar et~al.(2019)Ghoshdastidar, Gutzeit, Carpentier, and von
  Luxburg]{graph-comparison-other-8}
Debarghya Ghoshdastidar, Maurilio Gutzeit, Alexandra Carpentier, and Ulrike von
  Luxburg.
\newblock Two-sample hypothesis testing for inhomogeneous random graphs, 2019.
\newblock arXiv:1707.00833.

\bibitem[Ginestet et~al.(2017)Ginestet, Li, Balachandran, Rosenberg, and
  Kolaczyk]{graph-comparison-other-9}
Cedric~E Ginestet, Jun Li, Prakash Balachandran, Steven Rosenberg, and Eric~D.
  Kolaczyk.
\newblock {Hypothesis testing for network data in functional neuroimaging}.
\newblock \emph{Annals of Applied Statistics}, 11\penalty0 (2):\penalty0
  725--750, 2017.
\newblock ISSN 19417330.
\newblock \doi{10.1214/16-AOAS1015}.
\newblock URL
  \url{https://projecteuclid.org/download/pdfview{\_}1/euclid.aoas/1500537721}.

\bibitem[Agterberg et~al.(2020{\natexlab{a}})Agterberg, Tang, and
  Priebe]{agterberg2020nonparametric}
Joshua Agterberg, Minh Tang, and Carey Priebe.
\newblock Nonparametric two-sample hypothesis testing for random graphs with
  negative and repeated eigenvalues.
\newblock \emph{arXiv preprint arXiv:2012.09828}, 2020{\natexlab{a}}.

\bibitem[Shen and Vogelstein(2018)]{exact-equivalence-1}
Cencheng Shen and Joshua~T. Vogelstein.
\newblock The exact equivalence of distance and kernel methods for hypothesis
  testing, 2018.
\newblock arXiv:1806.05514.

\bibitem[Shen et~al.(2019{\natexlab{a}})Shen, Priebe, and
  Vogelstein]{exact-equivalence-2}
Cencheng Shen, Carey~E. Priebe, and Joshua~T. Vogelstein.
\newblock The exact equivalence of independence testing and two-sample testing,
  2019{\natexlab{a}}.
\newblock arXiv:1910.08883.

\bibitem[Vogelstein et~al.(2019)Vogelstein, Wang, Bridgeford, Priebe, Maggioni,
  and Shen]{mgc-0}
J.~T. Vogelstein, Q.~Wang, E.~Bridgeford, C.~E. Priebe, M.~Maggioni, and
  C.~Shen.
\newblock Discovering and deciphering relationships across disparate data
  modalities.
\newblock \emph{eLife}, 8:\penalty0 e41690, 2019.

\bibitem[Shen et~al.(2019{\natexlab{b}})Shen, Priebe, and Vogelstein]{mgc-1}
C.~Shen, C.~E. Priebe, and J.~T. Vogelstein.
\newblock From distance correlation to multiscale graph correlation.
\newblock \emph{Journal of the American Statistical Association},
  2019{\natexlab{b}}.

\bibitem[Lee et~al.(2019)Lee, Shen, Priebe, and Vogelstein]{mgc-2}
Y.~Lee, C.~Shen, C.~E. Priebe, and J.~T. Vogelstein.
\newblock Network dependence testing via diffusion maps and distance-based
  correlations.
\newblock \emph{Biometrika}, 2019.

\bibitem[Athreya et~al.(2018)Athreya, Fishkind, Tang, Priebe, Park, Vogelstein,
  Levin, Lyzinski, Qin, and Sussman]{athreya2018rdpg}
Avanti Athreya, Donniell~E. Fishkind, Minh Tang, Carey~E. Priebe, Youngser
  Park, Joshua~T. Vogelstein, Keith Levin, Vince Lyzinski, Yichen Qin, and
  Daniel~L Sussman.
\newblock Statistical inference on random dot product graphs: a survey.
\newblock \emph{Journal of Machine Learning Research}, 18\penalty0
  (226):\penalty0 1--92, 2018.
\newblock URL \url{http://jmlr.org/papers/v18/17-448.html}.

\bibitem[Sussman et~al.(2012)Sussman, Tang, Fishkind, and
  Priebe]{ase-consistency-1}
Daniel~L. Sussman, Minh Tang, Donniell~E. Fishkind, and Carey~E. Priebe.
\newblock A consistent adjacency spectral embedding for stochastic blockmodel
  graphs.
\newblock \emph{Journal of the American Statistical Association}, 107\penalty0
  (499):\penalty0 1119--1128, 2012.
\newblock \doi{10.1080/01621459.2012.699795}.
\newblock URL \url{https://doi.org/10.1080/01621459.2012.699795}.

\bibitem[Sussman et~al.(2014)Sussman, Tang, and Priebe]{ase-consistency-2}
Daniel~L. Sussman, Minh Tang, and Carey~E. Priebe.
\newblock Consistent latent position estimation and vertex classification for
  random dot product graphs.
\newblock \emph{IEEE Transactions on Pattern Analysis and Machine
  Intelligence}, 36:\penalty0 48--57, 2014.
\newblock \doi{10.1109/TPAMI.2013.135}.

\bibitem[Lyzinski et~al.(2014)Lyzinski, Sussman, Tang, Athreya, and
  Priebe]{ase-consistency-3}
Vince Lyzinski, Daniel~L. Sussman, Minh Tang, Avanti Athreya, and Carey~E.
  Priebe.
\newblock Perfect clustering for stochastic blockmodel graphs via adjacency
  spectral embedding.
\newblock \emph{Electronic Journal of Statistics}, 8\penalty0 (2):\penalty0
  2905--2922, 2014.
\newblock \doi{10.1214/14-EJS978}.
\newblock URL \url{https://doi.org/10.1214/14-EJS978}.

\bibitem[Zhu and Ghodsi(2006)]{zhu2006automatic}
Mu~Zhu and Ali Ghodsi.
\newblock Automatic dimensionality selection from the scree plot via the use of
  profile likelihood.
\newblock \emph{Computational Statistics \& Data Analysis}, 51\penalty0
  (2):\penalty0 918--930, 2006.

\bibitem[Athreya et~al.(2015)Athreya, Priebe, Tang, Lyzinski, Marchette, and
  Sussman]{Athreya2015}
A~Athreya, C~E Priebe, M~Tang, V~Lyzinski, D~J Marchette, and D~L Sussman.
\newblock {A Limit Theorem for Scaled Eigenvectors of Random Dot Product
  Graphs}.
\newblock Technical report, 2015.

\bibitem[Alyakin et~al.(2020)Alyakin, Agterberg, Helm, and
  Priebe]{correcting-nonpar}
Anton~A. Alyakin, Joshua Agterberg, Hayden~S. Helm, and Carey~E. Priebe.
\newblock Correcting a nonparametric two-sample graph hypothesis test for
  graphs with different numbers of vertices, 2020.

\bibitem[Agterberg et~al.(2020{\natexlab{b}})Agterberg, Tang, and
  Priebe]{on-two-sources}
Joshua Agterberg, Minh Tang, and Carey~E. Priebe.
\newblock On two distinct sources of nonidentifiability in latent position
  random graph models, 2020{\natexlab{b}}.
\newblock arXiv:2003.14250.

\bibitem[Sch{\"o}nemann(1966)]{schonemann1966generalized}
Peter~H Sch{\"o}nemann.
\newblock A generalized solution of the orthogonal procrustes problem.
\newblock \emph{Psychometrika}, 31\penalty0 (1):\penalty0 1--10, 1966.

\bibitem[Alvarez-Melis et~al.(2019)Alvarez-Melis, Jegelka, and
  Jaakkola]{alvarez2019towards}
David Alvarez-Melis, Stefanie Jegelka, and Tommi~S Jaakkola.
\newblock Towards optimal transport with global invariances.
\newblock pages 1870--1879, 2019.

\bibitem[Szekely and Rizzo(2014)]{szekely2014dcorr}
G.~Szekely and M.~Rizzo.
\newblock Partial distance correlation with methods for dissimilarities.
\newblock \emph{Annals of Statistics}, 42\penalty0 (6):\penalty0 2382--2412,
  2014.

\bibitem[Shen and Vogelstein(2019)]{shen2019chi}
Cencheng Shen and Joshua~T Vogelstein.
\newblock The chi-square test of distance correlation.
\newblock \emph{arXiv preprint arXiv:1912.12150}, 2019.

\bibitem[Eichler et~al.(2017)Eichler, Li, Litwin-Kumar, Park, Andrade,
  Schneider-Mizell, Saumweber, Huser, Eschbach, Gerber,
  et~al.]{eichler2017complete}
Katharina Eichler, Feng Li, Ashok Litwin-Kumar, Youngser Park, Ingrid Andrade,
  Casey~M Schneider-Mizell, Timo Saumweber, Annina Huser, Claire Eschbach,
  Bertram Gerber, et~al.
\newblock The complete connectome of a learning and memory centre in an insect
  brain.
\newblock \emph{Nature}, 548\penalty0 (7666):\penalty0 175, 2017.

\bibitem[Chung et~al.(2019)Chung, Pedigo, Bridgeford, Varjavand, Helm, and
  Vogelstein]{chung2019graspy}
Jaewon Chung, Benjamin~D. Pedigo, Eric~W. Bridgeford, Bijan~K. Varjavand,
  Hayden~S. Helm, and Joshua~T. Vogelstein.
\newblock Graspy: Graph statistics in python.
\newblock \emph{Journal of Machine Learning Research}, 20\penalty0
  (158):\penalty0 1--7, 2019.
\newblock URL \url{http://jmlr.org/papers/v20/19-490.html}.

\bibitem[Panda et~al.(2020)Panda, Palaniappan, Xiong, Bridgeford, Mehta, Shen,
  and Vogelstein]{panda2020hyppo}
Sambit Panda, Satish Palaniappan, Junhao Xiong, Eric~W. Bridgeford, Ronak
  Mehta, Cencheng Shen, and Joshua~T. Vogelstein.
\newblock hyppo: A comprehensive multivariate hypothesis testing python
  package.
\newblock 2020.

\end{thebibliography}

\end{document}